\documentclass[
 reprint,
superscriptaddress,
amsmath,
amssymb,
aps,
reprint,
superscriptaddress
]{revtex4-1}

\usepackage[T1]{fontenc}

\usepackage{graphicx}
\usepackage{dcolumn}
\usepackage{bm}
\usepackage{braket}
\usepackage{mathptmx}
\usepackage{color}
\usepackage{xcolor}
\usepackage[normalem]{ulem}

\begin{document}
\title{All-optical long-distance quantum communication\\with Gottesman-Kitaev-Preskill qubits}

\author{Kosuke Fukui}
\affiliation{Department of Applied Physics, School of Engineering, The University of Tokyo,\\7-3-1 Hongo, Bunkyo-ku, Tokyo 113-8656, Japan}
\author{Rafael N. Alexander}
\affiliation{Center for Quantum Information and Control, Department of Physics and Astronomy,\\University of New Mexico, Albuquerque, NM 87131, USA}
\affiliation{School of Science, RMIT University, Melbourne, Victoria 3000, Australia}
\author{Peter van Loock}
\affiliation{Institute of Physics, Johannes Gutenberg-Universit\"{a}t Mainz, Staudingerweg 7, 55128 Mainz, Germany}

\begin{abstract}
Quantum repeaters are a promising platform for realizing long-distance quantum communication and thus could form the backbone of a secure quantum internet, a scalable quantum network, or a distributed quantum computer.
Repeater protocols that encode information in single- or multi photon states are limited by transmission losses and the cost of implementing entangling gates or Bell measurements.
In this work, we consider implementing a quantum repeater protocol using Gottesman-Kitaev-Preskill (GKP) qubits. These qubits are natural elements for quantum repeater protocols, because they allow for deterministic Gaussian entangling operations and Bell measurements, which can be implemented at room temperature. The GKP encoding is also capable of correcting small displacement errors. At the cost of additional Gaussian noise, photon loss can be converted into a random displacement error channel by applying a phase-insensitive amplifier. Here we show that a similar conversion can be achieved in two-way repeater protocols by using phase-sensitive amplification applied in the postprocessing of the measurement data, resulting in less overall Gaussian noise per (sufficiently short) repeater segment.
We also investigate concatenating the GKP code with higher level qubit codes while leveraging analog syndrome data, post-selection, and path-selection techniques to boost the rate of communication. We compute the secure key rates and find that GKP repeaters can achieve a comparative performance relative to methods based on photonic qubits while using orders-of-magnitude fewer qubits.
\end{abstract} 

\maketitle

\section{Introduction}\label{Intro}
Reliable quantum communication protocols are an essential ingredient for creating a secure quantum internet~\cite{kimble2008quantum,wehner2018quantum}, implementing secure classical communication~\cite{bennett1984proceedings,ekert1991quantum, gisin2002quantum,scarani2009security,pirandola2019advances} and distributed quantum cryptographic protocols~\cite{bennett1993teleporting, mattle1996dense}.
Nowadays, quantum communication is routinely demonstrated over long distances ($>100$ km) using qubits encoded in the modal occupancy state of a single photon~\cite{ursin2007entanglement, ma2012quantum}. Such qubits are prone to being entirely lost into the environment (known as a photon loss)---the probability of being successfully transmitted (also referred to as the efficiency, $\eta$) decays exponentially with the transmission distance. By breaking long transmission distances into smaller, more manageable pieces, quantum repeaters~\cite{briegel1998quantum} overcome the exponentially bad scaling due to photon loss. A quantum repeater protocol implemented between two parties---the sender Alice and the receiver Bob---achieves a polynomial scaling of the efficiency's decay with the total distance between Alice and Bob.

The majority of quantum repeater protocols are based on quantum information encoded in simple superpositions of Fock states, typically just using single photons---though various proposals involve encoding information in coherent states of light~\cite{van2006hybrid, ladd2006hybrid}. These protocols can involve matter qubits
for storage or processing~\cite{yuan2008experimental, munro2012quantum, muralidharan2014ultrafast}, 
and can be memory free (and hence even ``all photonic''~\cite{azuma2015all,ewert2016ultrafast,lee2019fundamental}). In either case, the inherently probabilistic nature of practical entangling gates and/or Bell measurements for such qubits creates a bottleneck to engineering scalable repeater networks~\cite{calsamiglia2001maximum}. One strategy is to employ redundancy in the repeater protocol, which has the added benefit of providing a degree of robustness against photon loss~\cite{azuma2015all, ewert2016ultrafast,lee2019fundamental}.
More generally, by suppressing the effect of photon loss in the transmission channels through suitable quantum error correction codes and directly sending encoded logical qubits, the necessity of two-way classical communication (as required in memory-based repeaters to inform each station on successful entanglement distributions and manipulations of other stations) can be entirely circumvented. As a consequence, such so-called third-generation quantum repeaters~\cite{muralidharan2014ultrafast} are limited only by the elementary time units needed for locally preparing, processing, and detecting quantum states at each station, independent of distance-dependent waiting times for classical signals.
In principle, this approach of quantum communication allows for rates that match those in classical communication. 

The impact of photon loss and the cost of entangling gates or Bell measurements depend greatly on how we encode qubits into the bosonic modes. In this article, we propose a repeater protocol that employs the Gottesman-Kitaev-Preskill (GKP) qubit encoding~\cite{gottesman2001encoding}. This code allows for deterministic entangling gates and Bell measurements, both implementable at room temperature. In fact, the only nonlocal operation required for both GKP Bell-state generation and Bell measurement is a balanced beamsplitter interaction~\cite{walshe2020continuous}. The largest obstacle facing this approach is the difficulty of generating GKP qubits at optical frequencies. Individual GKP qubit states are non-Gaussian---they enable universal quantum computation with Gaussian only resources~\cite{baragiola2019all, yamasaki2020cost}. By leveraging strong nonlinearities available to trapped ion and superconducting systems, phononic and microwave mode GKP states have already been generated~\cite{fluhmann2019encoding, campagne2020quantum}. We note that conversion of microwave GKP states to optical frequencies may be possible in the future via microwave to optical transducers~\cite{andrews2014bidirectional, higginbotham2018harnessing}.
All-optical approaches to GKP qubit generation are also increasingly viable, particularly those that use photon counting as the source of non-Gaussianity~\cite{travaglione2002preparing, pirandola2004constructing, vasconcelos2010all, su2019conversion, eaton2019non}.

The GKP encoding is well-suited to protecting encoded information against Gaussian-random displacement noise~\cite{fukui2017analog}. This feature can be adapted to deal with photon loss by sending the GKP state through a phase-insensitive amplifier, either before or after the loss channel~\cite{albert2018performance}. The combined effect of the loss and amplification channels is an unknown Gaussian-random displacement, which can be corrected by using the standard GKP error correction procedure~\cite{gottesman2001encoding}. This succeeds with high probability for low levels of loss. However, phase insensitive amplification necessarily introduces extra noise into the system~\cite{haus1962quantum,caves1982quantum}, and so this strategy for combating photon loss is unlikely to be optimal.

In this work, we propose and compare multiple variant repeater protocols that use GKP qubits. In any repeater protocol, the transmission interval between Alice and Bob is divided into smaller segments with repeater stations placed at end points of each segment. GKP-repeater protocols may involve distributing GKP-encoded entangled states between adjacent repeater stations by transmission. Then information is sent from Alice to Bob by sequential teleportation from station to station, along with classical communication. We consider both one- and two-way repeater protocols. In the former case, both quantum states and classical information are transmitted in a single direction (towards Bob), from one repeater station to the next. In the latter case, sender stations that generate entangled states of GKP qubits are placed at the midpoints between repeater stations and transmit half of each entangled state to the neighboring repeater stations. Quantum states are then transmitted by teleportation, which requires one-way classical communication. It is therefore worth emphasizing that our error-correction-based variants of two-way protocols are distinct from memory-based quantum repeaters, where, as mentioned above, the two-way classical communication needed for reliable entanglement distribution puts high demands on the matter qubit's coherence times and fundamentally limits the achievable rates. Conceptually, our two-way schemes are similar to the all-photonic repeater protocol of Ref.~\cite{azuma2015all}. Practically, unlike that scheme ~\cite{azuma2015all} and also unlike the schemes of Refs.~\cite{ewert2016ultrafast,lee2019fundamental}, the optical measurements in our protocols are efficient homodyne detections throughout (except for the possible use of photon counting in GKP qubit generation), and also a much smaller number of qubits is needed in our case since GKP qubits are protected against errors from the start as opposed to single-photon qubits~\cite{azuma2015all,ewert2016ultrafast,lee2019fundamental}. Thus, our schemes inherit all benefits known from third-generation, all-optical quantum repeaters, with the additional advantage of ``hardware efficiency'', but the extra complication of GKP state generation (while state generation is also the most difficult part of the other all-optical proposals~\cite{azuma2015all,ewert2016ultrafast,lee2019fundamental}.)

In our two-way protocols, all modes experience the same amount of photon loss. In this case, we show that photon loss can be converted to Gaussian displacement noise via phase-sensitive amplifiers, circumventing the additional noise added by the aforementioned method that relies on phase-insensitive amplification. We note that for short-range point-to-point QKD with GKP qubits (without repeaters), both phase-sensitive and phase-insensitive amplification were considered in Ref.~\cite{gottesman2003secure}. In that setting, measurements were destructive and so did not have to preserve the codespace of the qubit. In this work, relying on the concept of teleportation-based syndrome detection, we describe how phase-sensitive amplifiers can be used to compensate against loss by nondestructive stabilizer measurement and while preserving the codespace. Moreover, this phase-sensitive amplification can be applied in the postprocessing of the measurement data, thus completely avoiding the need for physical squeezing gates at any stage of the protocol.

In addition to having information solely protected by the GKP encoding, we consider concatenation with higher level qubit codes to provide a greater degree of redundancy, and hence, robustness to error. GKP error correction---which is implemented through the teleportation step of our repeater protocol---produces analog syndrome data that can be used to improve decoding of the bare qubits and of the higher-level qubit code (if used)~\cite{fukui2017analog}. We propose to concatenate the GKP code with the qubit code proposed by Varnava, Browne, and Rudolph in Ref.~\cite{varnava2006loss}, which can tolerate up to $50\%$ lost qubits. 
 Our decoder estimates the quality of each transmitted GKP qubit from the analog syndrome data, and treats instances that result in low-quality states as lost qubits. These lost qubits are then dealt with by the Varnava code via a path-selection algorithm.

 Our numerical results compare one- and two-way protocols without higher level error correction, and also analyze the two-way implementation of the Varnava code. We show that our protocols can achieve a higher secure key rate compared with similar quantum repeater protocols based on photonic qubits~\cite{munro2012quantum,azuma2015all}. 

The rest of the paper is organized as follows. In Sec. \ref{Sec2}, we review the GKP qubit, our error model, single-qubit quantum error correction, postselected syndrome measurements, and amplification techniques that convert photon loss into a simpler Gaussian shift noise channel.
In Sec. \ref{Sec3}, we describe a quantum repeater protocol that uses bare GKP qubits (without a higher level code).
In Sec. \ref{Sec4}, we investigate  the quantum repeater protocols with GKP qubits concatenated with a higher quantum error correcting code, showing numerical calculations of the secure key rate.
Sec.\ref{conc} is devoted to discussion and conclusion. 

\section{Background}\label{Sec2}
In this section, we review the GKP qubits, the photon-loss channel, and the single-qubit level quantum error correction (SQEC).
We then describe two techniques used to improve quantum communication with the GKP encoding: 1) the highly-reliable measurement, which decreases the error probabilities via postselection, and 2) the phase-sensitive amplification of homodyne measurements.  

\subsection{The GKP qubit}
In this article, we work in units where $\hbar=1$ and the vacuum variances are $\langle\hat{q}^2\rangle_\text{vac}=\langle\hat{p}^2\rangle_\text{vac}=1/2$, where $\hat{a} = (\hat{q}+i\hat{p})/\sqrt{2}$. The logical states of the square-lattice GKP qubit are composed of a series of Gaussian peaks of width $\Delta$ contained in a larger Gaussian envelope of width 1/$\kappa$, with each peak separated by $\sqrt{\pi}$. In the position basis, the logical states $\ket {\widetilde{0}}$ and $\ket {\widetilde{1}}$ are given by
\begin{eqnarray}
\ket {\widetilde{0}} &\propto &  \sum_{t=- \infty}^{\infty} \int \mathrm{e}^{-2\pi\kappa^2t^2}\mathrm{e}^{-\frac{(s-2t\sqrt{\pi})^2}{2\Delta^2}}\ket{s}_q  ds,    \\ 
\ket {\widetilde{1}} &\propto &  \sum_{t=- \infty}^{\infty} \int \mathrm{e}^{-\pi\kappa^2(2t+1)^2/2} 
\mathrm{e}^{-\frac{(q-(2t+1)\sqrt{\pi})^2}{2\Delta^2}}\ket{s}_q  ds .     
\end{eqnarray}

Although in case of infinite squeezing ($\Delta \rightarrow 0$, $\kappa \rightarrow 0$) the states become the perfect GKP qubits with Dirac-comb wavefunctions~\cite{gottesman2001encoding}, in the case of finite squeezing the imperfect qubits are not orthogonal, and there is a nonzero probability of misidentifying $\ket {\widetilde{0}}$ with $\ket {\widetilde{1}}$, and vice versa. We choose $\kappa$ and $\Delta$ so that the variance of each tooth in the position and momentum observables is equal to $\sigma^{2}$, i.e., $\Delta^{2} = \kappa^{2} = 2\sigma^{2}$.

We will take finite-squeezing effects into account via a simplified model used in Ref.~\cite{menicucci2014fault} that neglects the effect of the phase-space envelopes. 
Nonideal GKP states are modeled by applying the following Gaussian bosonic completely positive trace-preserving (CPTP) map to ideal ($\sigma\mapsto 0$) code states $\ket{\bar{0}}$ and $\ket{\bar{1}}$: 
\begin{align}
	\mathcal{E}_{\sigma^{2}}(\rho) = \frac{1}{2\pi\sigma^{2} }\int \mathrm{d}s\mathrm{d}t e^{-(s^{2} + t^{2})/2\sigma^{2}}  e^{-i s \hat{p}} e^{i t \hat{q}} \rho e^{-i t \hat{q}} e^{i s \hat{p}} \label{eq:gauschan}
\end{align}
Notice that the outputs of this channel are mixed states, even when $\rho$ was pure.
This channel adds $\sigma^{2}$ to the variance of the initial state $\rho$ in both the position and the momentum quadrature.

The probability of misidentifying the bit value of a nonideal GKP qubit,  $P_{\rm fail}(\sigma^2)$, is approximately
\begin{equation}
P_{\rm fail}(\sigma^2) = 1-\int_{\frac{-\sqrt{\pi}}{2}}^{\frac{\sqrt{\pi}}{2}} dx \frac{1}{\sqrt{2\pi {\sigma} ^2}} {\rm exp}(-\frac{x^2}{2{\sigma} ^2}).
\label{eq3}
\end{equation}

\subsection{Photon-loss channel}
The dominant noise introduced by transmission is described by the photon-loss channel, ${\cal L}$. This can be modeled as an unwanted beamsplitter coupling with an environmental vacuum state.
In the Heisenberg picture, ${\cal L}$  transforms the $\hat{q}$ and $\hat{p}$ quadratures as
\begin{equation}
\hat{q} \to   \sqrt{\eta}  \hat{q}+\sqrt{1-\eta}\hat{q}_{\rm vac}, \hspace{10pt} \hat{p} \to  \sqrt{\eta}  \hat{p}+\sqrt{1-\eta}\hat{p}_{\rm vac}, 
\end{equation}
where $\sqrt{\eta}$ is the transmittance coefficient (square root of the efficiency, $\eta$), and $\hat{q}_{\rm vac}(\hat{p}_{\rm vac})$ is the position (momentum) quadrature of the vacuum state.
After the loss channel, the variances in the $\hat{q}$ and $\hat{p}$ quadratures transform as
\begin{eqnarray}
{\sigma^{2}} _{{\rm in},q }   \to   {\eta}\sigma^{2}_{{\rm in},q} +(1-{\eta})/2 , \\
{\sigma^{2}} _{{\rm in},p }   \to   {\eta}\sigma^{2}_{{\rm in},p} +(1-{\eta})/2 , 
\end{eqnarray}
where ${\sigma^{2}} _{{\rm in},q }({\sigma^{2}} _{{\rm in},p})$ are the variances of the GKP qubits before the transmission loss channel ${\cal L}$.
In this work, we assume that the ${\sigma^{2}} _{{\rm in},q }$ and ${\sigma^{2}} _{{\rm in},p}$ of the single GKP qubit are equal to $\frac{1}{2}{\rm e}^{-2r}$, where $r$ is the squeezing parameter. The parameter $\eta$ is related to the distance of the transmission $L$ (km) as $\eta={\rm exp}(-L/L_{\rm att})$ with the attenuation length $L_{\rm att} = 22$ km.

The conjugate to photon loss is the amplification channel, i.e., 
\begin{align}
	\text{tr}\left[ \hat{\sigma} \mathcal{L}(\hat{\rho}) \right] =\text{tr}\left[ \mathcal{A}(\hat{\sigma}) \hat{\rho} \right]
\end{align}	
for any density matrices $\hat{\sigma}$ and $\hat{\rho}$, where in the Heisenberg picture, ${\cal A}$  transforms the $\hat{q}$ and $\hat{p}$ quadratures as
\begin{equation}
	\hat{q} \to   \sqrt{\frac{1}{\eta}}  \hat{q}+\sqrt{\frac{1}{\eta}-1}\hat{q}_{\rm vac}, \hspace{10pt} \hat{p} \to  \sqrt{\frac{1}{\eta}}  \hat{p}+\sqrt{\frac{1}{\eta}-1}\hat{p}_{\rm vac}, 
\end{equation}
and transforms the variances in the $\hat{q}$ and $\hat{p}$ quadratures by
\begin{eqnarray}
	{\sigma^{2}} _{{\rm in},q }   \to \frac{1}{\eta}\sigma^{2}_{{\rm in},q} +\frac{1-\eta}{2\eta}, \\
	{\sigma^{2}} _{{\rm in},p }   \to \frac{1}{\eta}\sigma^{2}_{{\rm in},p} +\frac{1-\eta}{2\eta}. 
\end{eqnarray}

\subsection{Single-qubit level QEC with transmission losses}
If a GKP qubit is displaced by an amount less than half the lattice spacing, this perturbation can be diagnosed by measuring position and momentum mod $\sqrt{\pi}$.
This displacement can then be undone by an active displacement, or by updating the ``quadrature frame" of phase space. Relevant for Gaussian noise with variance $\sigma^{2}$ is the probability of a displacement occurring that is greater than half the lattice spacing, as described in Eq.~(\ref{eq3}). In this case, an undetectable logical Pauli error is said to have occured. The above description assumes perfect error correction, and does not take into account additional noise that arises in the measurement of the GKP stabilizers. 

Here we compute the probability of logical error for the teleportation-style GKP error correction introduced in Ref.~\cite{walshe2020continuous}. This approach comes with two advantages: 1) the only nonlocal operations required are beamsplitters, and so this method completely eschews the squeezing operations required in other methods, e.g., that of Ref.~\cite{glancy2006error}, and 2) the approximate error correction channel can be decomposed into an ideal error correction operation sandwiched between two Gaussian noise channels. 

Before describing teleportation-based error correction, we review some definitions and facts.
Define the so-called qunaught state
\begin{align}
     \ket{\varnothing} \propto \sum_{k=-\infty}^{\infty} e^{-i\sqrt{2\pi}k\hat{p}} \ket{0}_{q}
     = \sum_{k=-\infty}^{\infty} e^{i \sqrt{2\pi}k\hat{q}} \ket{0}_{p}
 \end{align}
which is useful for quantum sensing applications~\cite{duivenvoorden2017single}. This state has $\sqrt{2\pi}$ periodicity in both position and momentum. Sending two qunaught states through a 50:50 beamsplitter generates a GKP Bell-state~\cite{walshe2020continuous}
 \begin{align}
     B  \ket{\varnothing}\otimes  \ket{\varnothing} = \frac{\ket{\bar{0}\bar{0}}+\ket{\bar{1}\bar{1}}}{\sqrt{2}} \label{eq:LObell}.
 \end{align}
It is straightforward to verify that any beamsplitter commutes with a pair of equal-strength Gaussian noise channels:
\begin{align}
    \mathcal{B} \circ \left[ \mathcal{E}_{\sigma^{2}}\otimes \mathcal{E}_{\sigma^{2}} \right] = \left[ \mathcal{E}_{\sigma^{2}}\otimes \mathcal{E}_{\sigma^{2}} \right] \circ  \mathcal{B}, \label{eq:Ecomm}
\end{align}
where $\mathcal{B}(\rho_a\otimes\rho_b) = B \rho_a\otimes\rho_b B^{\dagger}$ is the superoperator form of $B$. 
This property also holds for the photon-loss channel
\begin{align}
	\mathcal{B} \circ \left[ \mathcal{L}\otimes \mathcal{L} \right] = \left[ \mathcal{L}\otimes \mathcal{L} \right] \circ  \mathcal{B}. \label{eq:Lcomm}
\end{align}

A beamsplitter followed by a pair of homodyne detectors is equivalent to a displacement on the first mode followed by a projector onto an EPR pair:
\begin{align}
    {_{q}\left\langle\frac{m_1}{\sqrt{2}}\right\vert}{_p\left\langle\frac{m_2}{\sqrt{2}}\right\vert} B &=  {_{q}\bra{0}}{_p\bra{0}} B D_1(\mu) \nonumber \\ &=  {_{p}\bra{0}}{_q\bra{0}} C_X D_1(\mu) \nonumber\\
    &= {_{p}\bra{m_2}}{_q\bra{m_1}} C_X, \label{eq:bellproj}
\end{align}
where $D$ is the displacement operator, $\mu = -(m_1 +i m_2)/\sqrt{2}$, and $C_X = e^{-i \hat{q}_1\hat{p}_2}$.

\begin{figure}
    \centering
    \includegraphics[width=\linewidth]{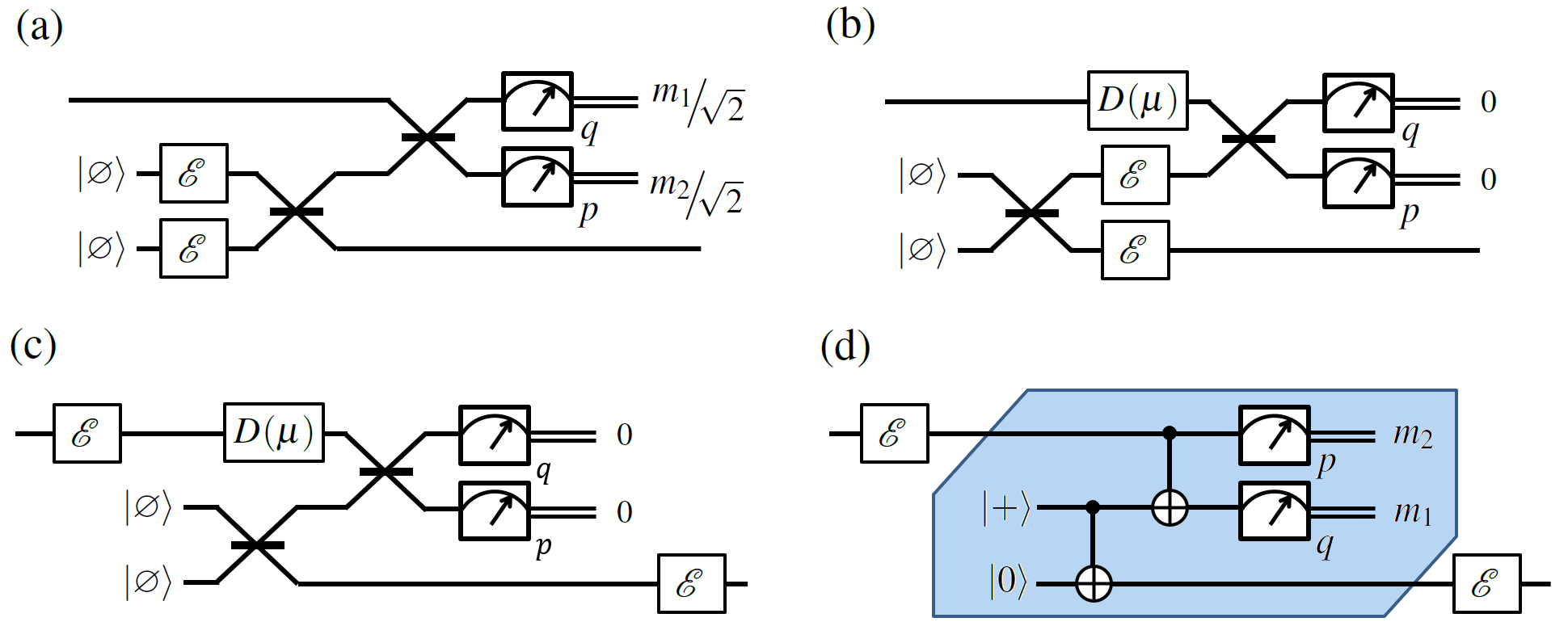}
    \caption{\textbf{(a)} Teleportation-based error correction as formulated in Ref.~\cite{walshe2020continuous}, using the Gaussian noise approximation of finite energy effects, Eq.~(\ref{eq:gauschan}). \textbf{(b)} We can bring the noise maps $\mathcal{E}$ through the beamsplitter as described in Eq.~(\ref{eq:Ecomm}). In addition, the CV Bell measurement is equivalent to a displacement plus a projection onto the EPR pair, as shown in Eq.~(\ref{eq:bellproj}). \textbf{(c)} One noise map $\mathcal{E}$ can be ``bounced'' from the second circuit wire to the first using the EPR pair. This implements a transpose map, which can be ignored as $\mathcal{E}^{\text{T}}=\mathcal{E}$. The channel is then pulled to the left of the displacement $D(\mu)$ (they commute). \textbf{(d)} Two grid states sent through a beamsplitter are equivalent to a GKP Bell pair~\cite{walshe2020continuous}. The beamsplitter in the EPR projection can be replaced with a $C_X$ gate. The displacement $D(\mu)$ can be re-combined with the homodyne detectors using Eq.~(\ref{eq:bellproj}). The circuit in the blue region implements perfect GKP teleportation and is equivalent to GKP error correction with modular position and momentum outcomes $m_1$ and $m_2$ respectively.   }
    \label{fig:GKPLOformulation}
\end{figure}

We can use these facts to understand the linear optics implementation of GKP error correction, as shown in Fig.~\ref{fig:GKPLOformulation}.
This reveals that SQEC takes a state with input variances in $q$ and $p$, denoted $\sigma^{2}_{q, \text{in}}$ and $\sigma^{2}_{p, \text{in}}$, then adds variance from the ancilla
\begin{align}
    \sigma^{2}_{q, \text{in}} \mapsto \sigma^{2}_{q, \text{in}} + \sigma^{2}, \\
    \sigma^{2}_{p, \text{in}} \mapsto \sigma^{2}_{p, \text{in}} + \sigma^{2},
\end{align}
then we implement perfect GKP error correction, resulting in a logical $X$ or $Z$ error with probability $P_{\rm fail}(\sigma^{2}_{q, \text{in}} + \sigma^{2})$ and $P_{\rm fail}(\sigma^{2}_{p, \text{in}} + \sigma^{2})$, respectively. At this stage, the state is a perfect GKP state. Finally, the quality of the output state is ``reset'' to the starting quality, so that each tooth has variance
\begin{align}
    \sigma^{2}_{q, \text{out}} = \sigma^{2}, \\
    \sigma^{2}_{p, \text{out}} = \sigma^{2}. 
\end{align}

\begin{figure}[t]
\centering \includegraphics[angle=0, scale=1.0]{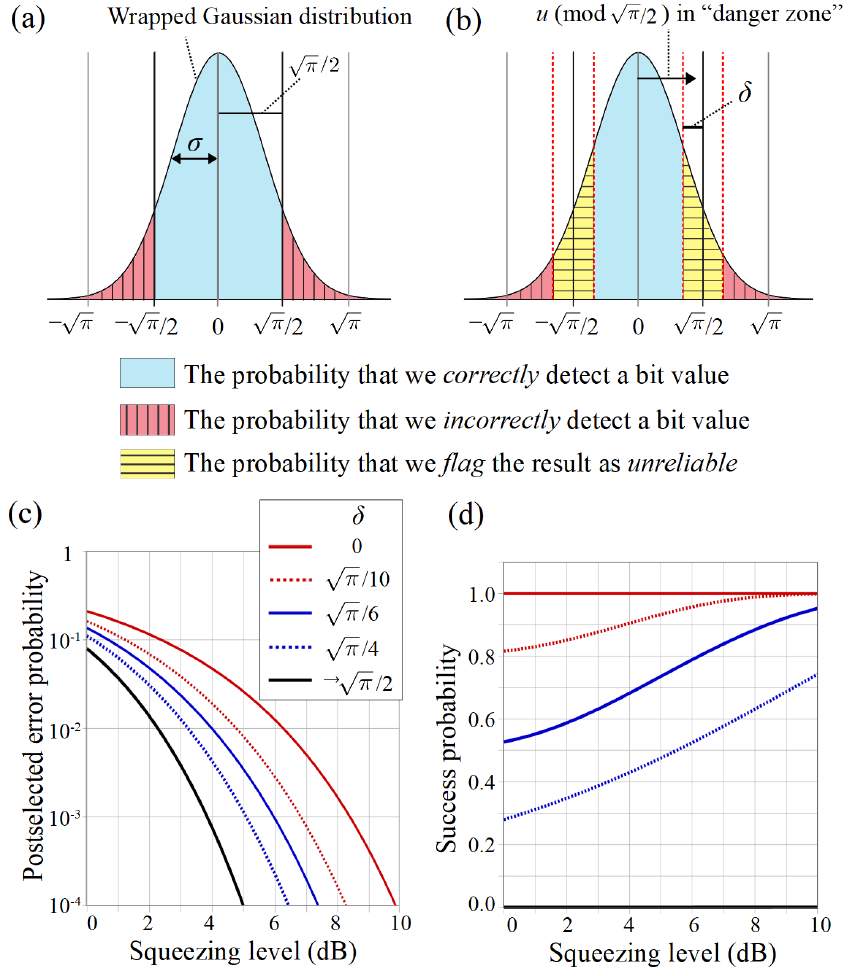} 
     \setlength\abovecaptionskip{10pt}
\caption{Introduction of the highly-reliable measurement (HRM). (a) The conventional measurement of the GKP qubit, where the Gaussian probability distribution following the deviation of the GKP qubit has variance $\sigma^{2}$. The plain blue region and the red region with vertical lines represent the different code words $(k-1)$ mod 2 and $(k+1)$ mod 2, respectively. The red regions marked with vertical lines correspond to the probability of incorrect decision of the bit value. (b) The highly-reliable measurement. One of the dotted lines represents an upper limit $v_{\rm up}$. The yellow areas with horizontal lines show the probability that the results of the measurement are discarded by introducing $v_{\rm up}$. The vertical line areas show the probability that our method fails. (c) The probability of misidentifying the bit value with the HRM, $E_{\delta}(\sigma^2)$. (d) The success probability of the HRM, $P_{\delta}^{\rm suc}$.
}
\label{fig2}
\end{figure}
\subsection{Highly-reliable measurement}\label{sec:HRM}
As was pointed out in Ref.~\cite{fukui2018high}, the probability of a logical error in GKP error correction depends on the continuous values of each syndrome measurement (rather than just the binned value). Said another way, conclusions drawn from syndrome measurements that fall closer to the ``midpoints'' between the GKP wavefunction teeth are less reliable. Discarding GKP qubits that result in untrustworthy syndromes improves the average fidelity of the rest. This is the key idea behind the so-called postselected highly-reliable measurement (HRM), which we now describe. 
 
Measuring GKP qubit stabilizers results in measurement values of the form $(2 t + k)\sqrt{\pi}+{\Delta}_{\rm m}$ $(t = 0, \pm 1, \pm 2,\cdots.)$, where $k$ is the identified bit value and ${\Delta}_{\rm m}$ is the measured deviation whose absolute value is smaller than $\sqrt{\pi}/2$.
The HRM can reduce the probability of misidentifying the bit value of the GKP qubit stabilizer by introducing a more restrictive cutoff margin than that given by the midpoints between lattice teeth. We parameterize this more restricted decision margin by $v_{\rm up}$, which is the cut-off line, and $\delta$, which is the margin of the unreliable outcome region. Binning without postselection sets an upper limit $v_{\rm up}$ and margin $\delta$ for $|\Delta_{m}|$ at $\sqrt{\pi}/2$ and 0, respectively, and assigns the bit value $k$ = $(2 t + k)\sqrt{\pi} $, as shown in Fig.~\ref{fig2}(a). 
In contrast, the HRM decision sets an upper limit at $v_{\rm up}(<\sqrt{\pi}/2)$ for the maximum deviation as shown in Fig.~\ref{fig2}(b). If the above condition $|\Delta_{\rm m}| < v_{\rm up}$ is not satisfied, we discard the result. Since a measurement error occurs when $|\bar{\Delta}|$ exceeds $|\sqrt{\pi}/2+\delta|$, the error probability decreases with increasing (decreasing) $\delta$ $(v_{\rm up})$. Using this method comes at the cost of reducing success probability of the measurement. The probability $P_{\delta}$ to obtain the correct bit value with the HRM is given by
\begin{align}
	P_{\delta}= \frac{{P_{\delta}^{\rm cor}}}{{P_{\delta}^{\rm cor}}+{P_{\delta}^{\rm in}}}, \label{eq:HRMprob}
\end{align}	
where ${P_{\delta}^{\rm cor}}$ is the probability that the true deviation $|\bar{\Delta}|$ falls in the correct area, and ${P_{\delta}^{\rm in}}$ is the probability that the true deviation $|\bar{\Delta}|$ falls in the incorrect area. ${P_{\delta}^{\rm cor}}$ and ${P_{\delta}^{\rm in}}$ for the GKP qubit of the variance $\sigma^2$ are given by
\begin{equation}
P_{\delta}^{\rm cor}= \sum_{k=-\infty}^{+\infty} \int_{2k\sqrt{\pi}-\frac{\sqrt{\pi}}{2}+\delta}^{2k\sqrt{\pi}+\frac{\sqrt{\pi}}{2}-\delta} dx \frac{1}{\sqrt{2\pi {\sigma}^2}}\mathrm{e}^{-\frac{x^2}{{2{\sigma}^2}}}
\end{equation}
and 
\begin{equation}
P_{\delta}^{\rm in}= \sum_{k=-\infty}^{+\infty} \int_{(2k+1)\sqrt{\pi}-\frac{\sqrt{\pi}}{2}+\delta}^{(2k+1)\sqrt{\pi}+\frac{\sqrt{\pi}}{2}-\delta} dx \frac{1}{\sqrt{2\pi {\sigma}^2}}\mathrm{e}^{-\frac{x^2}{{2{\sigma}^2}}},
\end{equation}
respectively. 
The probability of misidentifying the bit value with the HRM, $E_{\delta}(\sigma^2)$, is given by
\begin{align}
	E_{\delta}(\sigma^2)=1-P_{\delta} \label{eq:ehrm}
\end{align}	
	 for the qubit whose variance is $\sigma^2$.
Then, the success probability of the HRM, $P_{\delta}^{\rm suc}$, is given by 
\begin{equation}
P_{\delta}^{\rm suc}=P_{\delta}^{\rm cor}+P_{\delta}^{\rm in}. \label{eq:psuc}
\end{equation}
Figs.~\ref{fig2}(c) and (d) show probabilities $E_{\delta}(\sigma^2)$ and $P_{\delta}^{\rm suc}$, respectively, as a function of the squeezing level, where the squeezing level is equal to -10${\rm log}_{10}(2\sigma^2).$

\subsection{Amplification}
Photon loss has two effects on quadrature values: it rescales their amplitude, bringing them closer to the origin, and it introduces Gaussian noise. Linear amplification is a convenient technique for restoring the means of the quadrature values to their original locations. Phase-insensitive amplification can restore both quadratures simultaneously at the cost of additional Gaussian noise. Phase-sensitive amplification is equivalent to squeezing, and thus restores the amplitude of one quadrature at the expense of the other. Here, we review three methods for compensating for photon-loss via amplification. The first two involve phase-insensitive amplification, which can be applied before or after the loss channel. The third applies to a more specific case, where equal photon loss is applied before an EPR measurement. In this case, photon losses can be converted to Gaussian noise acting on a single input mode via phase-sensitive amplification. 

In the first two methods, amplification is implemented in the quantum repeater stations either after or before transmission. We refer to these methods as \emph{postamplification} and \emph{preamplification}, respectively. For postamplification, the quadrature variances in both $\hat{q}$ and $\hat{p}$ are transformed as
\begin{equation}
\sigma^2_{\rm in} \to   \sigma^2_{\rm in} + \frac{1-\eta}{\eta}, \label{eq:amp1}
\end{equation}
and for preamplification, the quadrature variances in both $\hat{q}$ and $\hat{p}$ are transformed as
\begin{equation}
	\sigma^2_{\rm in} \to   \sigma^2_{\rm in} + 1-\eta,  \label{eq:amp2}
\end{equation}
where $\sigma^2_{\rm in}$ is the initial variance before loss~\cite{noh2018quantum}. Note that the latter introduces less noise than the former.

\begin{figure}
\includegraphics[width=\linewidth]{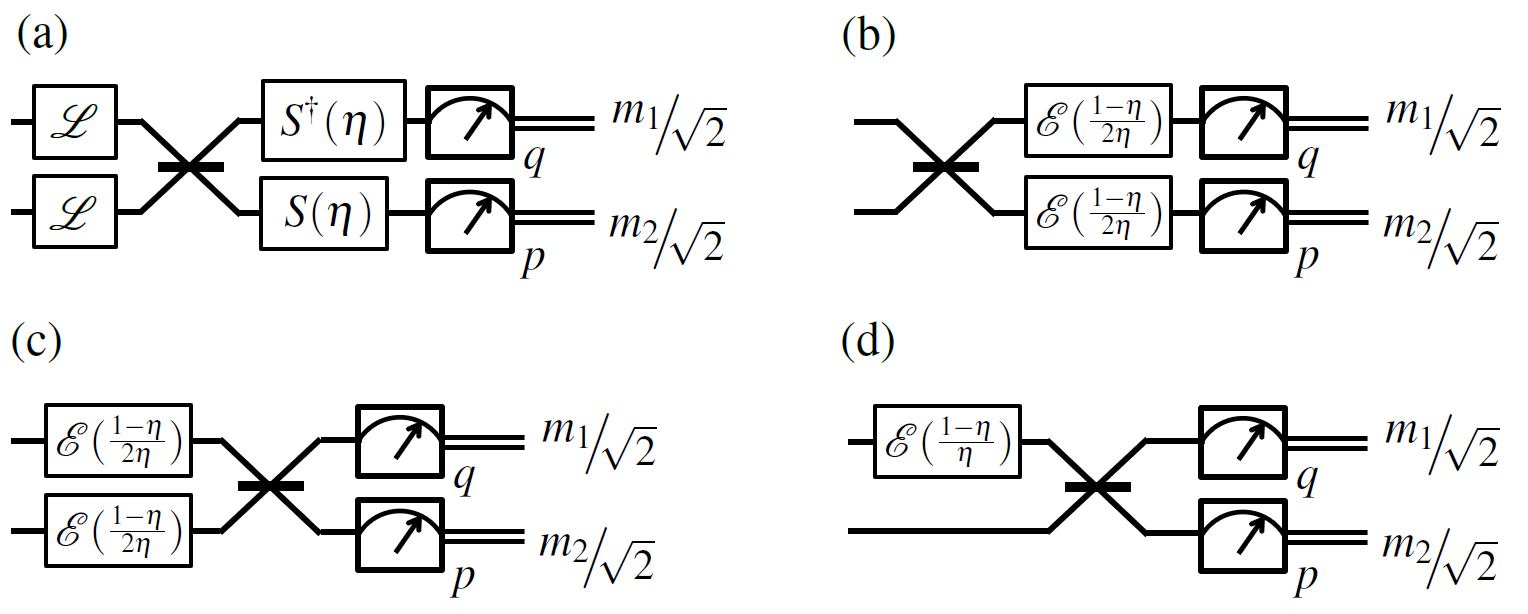}
\caption{ (a) Consider a pair of photon loss channels each with efficiency paramter $\eta$ that precede an EPR-basis measurement. Assume that the outcomes are rescaled in a way equivalent to the squeezing operators shown. Using Eq.~(\ref{eq:Lcomm}), the photon loss channels can be commuted with the 50:50 beamsplitter so that they act directly before the squeezed/rescaled detectors.  (b) A photon-loss channel and a phase-sensitive amplifier acting before a homodyne detector is equivalent to a Gaussian noise channel with variance $(1-\eta)/2\eta$, as described in Eqs.~(\ref{eq:sqampq}) and (\ref{eq:sqampp}). (c) A pair of equal strength additive Gaussian noise channels commute with a beamsplitter, Eq.~(\ref{eq:Ecomm}). (d) Equivalently, the additive noise channel on either input can be pushed onto the other input.}
\label{fig:ccampfig}
\end{figure}

The third technique consists of rescaling the homodyne outcomes on a classical computer after measuring in the EPR basis. More specifically, after we obtain the measurement value $m_q (m_p)$ in the $\hat{q}(\hat{p})$ quadrature, we multiply the value by $1/\eta$ on the classical computer. This is equivalent to applying a single-mode squeezer $S^{\dagger} (\eta)$ ($S(\eta)$) that rescales the position (momentum) quadrature by a factor of $1/\eta$. This will convert uniform photon loss into uniform Gaussian noise, as shown in Fig.~\ref{fig:ccampfig}.

A photon-loss channel with efficiency $\eta$ before a homodyne detector adds Gaussian noise, and rescales the outcome. This can be verified by considering the action of the adjoint channel on position and momentum eigenstates, i.e., 
\begin{align}
	\left[ \mathcal{A} \left(\left\vert{\frac{\eta m_1 }{\sqrt{2}}}\right\rangle_q\!\!{\left\langle{\frac{\eta m_1}{\sqrt{2}}}\right\vert}\right) \right]^{\dagger}
	=
		\left[ \mathcal{E}_{\frac{1-\eta}{2\eta}} \left(\left\vert{\frac{ m_1 }{\sqrt{2}}}\right\rangle_q\!\!{\left\langle{\frac{ m_1}{\sqrt{2}}}\right\vert}\right) \right]^{\dagger} \label{eq:sqampq}
\end{align}
and similarly for momentum eigenstates
\begin{align}
	\left[ \mathcal{A} \left(\left\vert{\frac{\eta m_2 }{\sqrt{2}}}\right\rangle_p\!\!{\left\langle{\frac{\eta m_2}{\sqrt{2}}}\right\vert}\right) \right]^{\dagger}
	=
	\left[ \mathcal{E}_{\frac{1-\eta}{2\eta}} \left(\left\vert{\frac{ m_2 }{\sqrt{2}}}\right\rangle_p\!\!{\left\langle{\frac{ m_2}{\sqrt{2}}}\right\vert}\right) \right]^{\dagger}. \label{eq:sqampp}
\end{align}

This technique converts a pair of loss channels into a pair of Gaussian noise channels that increase the variance by
\begin{equation}
	\sigma^2_{\rm in} \to   \sigma^2_{\rm in} + \frac{1-\eta}{2\eta},  \label{eq:amp3}
\end{equation}
or a single noise channel that increases the variance by
\begin{equation}
	\sigma^2_{\rm in} \to   \sigma^2_{\rm in} + \frac{1-\eta}{\eta}.  \label{eq:amp3alt}
\end{equation}
These are shown in Figs.~\ref{fig:ccampfig} (c) and (d), respectively. 

It is worth noting that the postamplification method can also be implemented in postprocessing of data on a classical computer (CC). To do so, one rescales the measurement outcomes and then adds them to a Gaussian random number. It is this last step that makes Eq.~(\ref{eq:amp3}) an improvement over Eq.~(\ref{eq:amp2}). For this reason, we refer to this method as \emph{CC-amplification}.

\begin{figure}[t]
\centering \includegraphics[angle=0, scale=0.9]{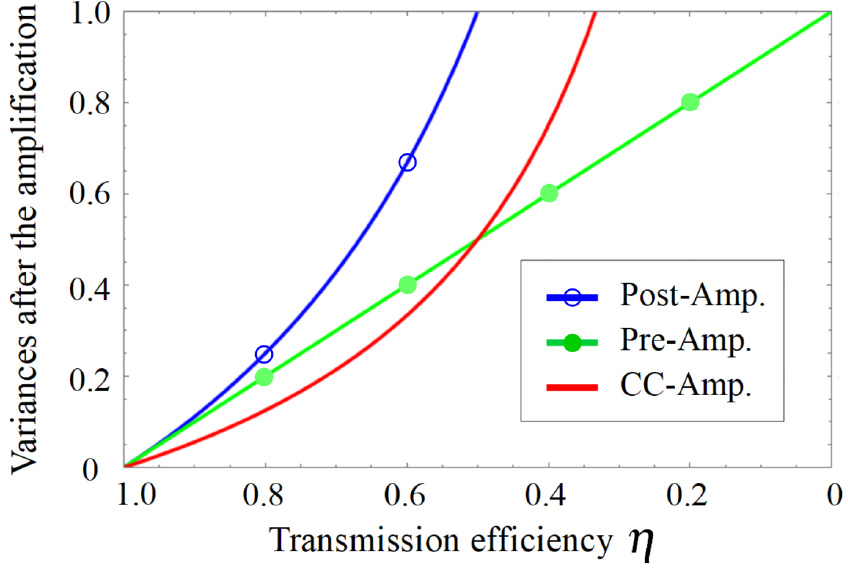} 
     \setlength\abovecaptionskip{10pt}
\caption{Three amplification techniques for point-to-point QKD, where the transmittance rate is $\eta$. Blue line with open circles, green line with filled circles, and red line depict the variances, $\frac{1-\eta}{\eta}$, ${1-\eta}$, and $\frac{1-\eta}{2\eta}$, respectively.
}
\label{fig4}
\end{figure}

Fig.~\ref{fig4} shows the variance values for the three amplification techniques, $\frac{1-\eta}{\eta}$, ${1-\eta}$, and $\frac{1-\eta}{2\eta}$, as a function of the transmittance $\eta$.
One can see that the amplification on the classical computer is better than the other types of amplification at point-to-point quantum key distribution (QKD)~\cite{bennett1984proceedings}.
In Secs. \ref{Sec3} and \ref{Sec4}, we apply these amplifications to quantum repeater protocolswithout or with an additional, higher-level quantum error correcting code, respectively, comparing the performance of these amplifiers.

In this work, while we focus exclusively on the pure-loss channel, our treatment can be straightforwardly extended to the thermal-loss channel, where we rescale the outcomes of the homodyne measurement and obtain a different amount of Gaussian noise. Although these extensions will generalize our method, the fundamental results are not expected to be changed.

\begin{figure*}[t]
\centering \includegraphics[angle=0, scale=2.0]{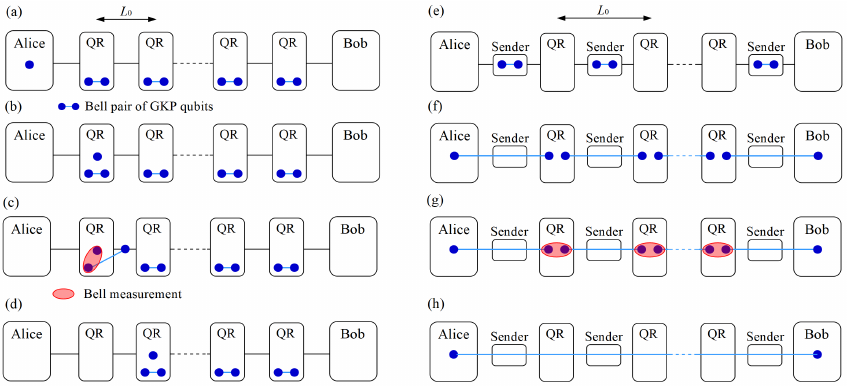}   
    \caption{A schematic drawing of the one-way ((a)-(d)) and two-way ((e)-(h)) quantum repeater protocols. (a) The sender (Alice) prepares the GKP qubit. Each of the nodes prepare a Bell pair of GKP qubits. (b) Alice sends the qubit to the first quantum repeater station. (c) The first quantum repeater station performs the SQEC protocol. After this, the state is sent to the next station. (d) Each quantum repeater station receives the state, performs the SQEC protocol, and then sends the state to the next station in sequence. Finally, Bob receives the qubit from the $N_{\rm QR}$-th quantum repeater.
In the SQEC with the HRM, any quantum communication attempt is aborted if the measurement outcome is greater than $v_{\rm up}$.   
(e) Each senders prepare a Bell pair. (f) Each sender except the first and  $N_{\rm QR+1}$-th sender sends a GKP Bell pair to their neighboring quantum repeater stations. The first and  $N_{\rm QR+1}$-th sender send half a Bell pair to Alice and Bob, respectively. 
(g) Implementation of the Bell measurements to perform SQECs.
(h) Generation of the entanglement between Alice and Bob. 
}
\label{fig5}
\end{figure*}

\section{Quantum repeater protocol without higher level encoding}\label{Sec3}
In this section, we describe the quantum repeater protocols for one- and two-way quantum communication without concatenation with any higher level quantum error correcting code. We calculate the secure-key rates for all three amplification techniques, and show the merit of using the HRM to improve the secure-key rate at the expense of the success probability (equivalently, the raw rate) of quantum communication.
Since the CC-amplification requires uniform levels of photon loss on both modes participating in the EPR measurement, this technique cannot be applied to one-way quantum repeater protocols. We also consider an additional variant of two-way repeater protocols where an extra round of quantum error correction is applied within each quantum repeater station. This protocol is compatible with both post- and preamplification techniques. Incidentally, we note that one- and two-way protocols are also commonly referred to as asymmetric and symmetric protocols, respectively.

\subsection{One-way quantum repeater protocol}
Figs.~\ref{fig5}(a)-(d) show the schematic view of the one-way repeater protocol used in this work.  $N_{\rm QR}$ is the number of quantum repeater stations between Alice and Bob. 
There are four steps in the protocol. 
In step 1, Alice prepares and sends the GKP qubit to the first quantum repeater, as shown in Fig.~\ref{fig5}(a). 
In the case of the preamplification, we perform the amplification before the transmission occurs. 
In step 2, the first repeater receives the GKP qubit (Fig.~\ref{fig5}(b)) and implements the SQEC protocol by using a Bell measurement. This corrects deviations that arise from both the intrinsic finite squeezing effects from the initial GKP qubit ($\mathcal{E}$), and the effects of photon loss ($\mathcal{L}$), as shown in Fig.~\ref{fig5}(c). 
In the case of postamplification, the repeater station performs amplification before the SQEC protocol.
After the SQEC protocol, the first repeater sends the qubit to the second repeater.
In step 3, the $i$-th ($i$=2, \dots, $N_{\rm QR}$) repeater receives the qubit and sends it to the $(i+1)$-th repeater after the SQECs in the same way as is described in step 2.
In the SQEC protocol with HRM, the quantum communication attempt is aborted if the measurement outcome gives more than $v_{\rm up}$.   
In step 4, Bob receives the qubit, can perform SQEC (or not), and then measures it, as shown in Fig.~\ref{fig5}(d). 

Now we consider the variances of the GKP qubits before each SQEC step of the repeater protocol.
In step 1, the initial variances in $q$ and $p$ of the GKP qubit Alice prepares are $({\sigma_{q}}^2,{\sigma_{p}}^2)=(\sigma^2,\sigma^2)$, respectively.
After both amplification and loss, the variances for post- and preamplification cases are $( \sigma^2 + \frac{1-\eta}{\eta},\sigma^2+ \frac{1-\eta}{\eta})$ and $(\sigma^2+ {1-\eta},\sigma^2+ {1-\eta})$, respectively. 
Performing SQEC introduces additional Gaussian noise from the finite energy ancilla, and so the variances before ideal GKP error correction become $( 2\sigma^2 + \frac{1-\eta}{\eta},2\sigma^2+ \frac{1-\eta}{\eta})$ and $(2\sigma^2+ {1-\eta},2\sigma^2+ {1-\eta})$, respectively. These variances follow from Eqs.~(\ref{eq:amp1}) and (\ref{eq:amp2}).
After SQEC, the variances of the GKP states are reset to $({\sigma_{q}}^2,{\sigma_{p}}^2)=(\sigma^2,\sigma^2)$ by the finite energy effects of the ancilla states. The process then repeats.

The secure key rate will be obtained by using the above variances in quantum repeaters as shown in the subsection III.~D.

The $X$ and $Z$ logical error rates are denoted $E^X$ and $E^Z$. These errors occur with equal and independent probabilities. For a single transmission step, $E^X$ and $E^Z$ equal 
\begin{align}
E_{\delta}\left(2\sigma^2 + \frac{1-\eta}{\eta}\right) \quad\quad &\text{one-way postamplification} \label{eq:owpoa}\\ 
E_{\delta}(2\sigma^2+ {1-\eta}) \quad\quad &\text{one-way preamplification}  \label{eq:owpra}
\end{align}

\subsection{Two-way quantum repeater protocol}
Now we describe the two-way quantum repeater protocol.
Figs.~\ref{fig5}(e)-(h) show the schematic view of the protocol.
In step 1, each of the sender nodes prepares a GKP Bell pair, then sends it to neighboring quantum repeater stations as shown in Fig.~\ref{fig5}(e). 
In the case of preamplification, amplification is performed on both qubits that make up the Bell pair before they are transmitted. Note that both modes experience the same amount of loss, but are only transmitted along half the distance $L_0/2$ relative to the one-way protocol. 
In step 2, each of the quantum repeater stations receives a GKP qubit from each neighboring sender node, as shown in Fig.~\ref{fig5}(f). 
In the case of postamplification, amplification occurs at the quantum repeater stations upon receiving the GKP qubits.

Then, the repeater stations implement Bell measurements via a beamsplitter coupling, implementing SQECs as shown in Fig.~\ref{fig5}(g). In the case of CC-amplification, amplification occurs by rescaling the homodyne outcomes by a factor of $\sqrt{\eta}$.
In step 3, Alice and Bob obtain the Bell pair after the feedforward operations if and only if there is no failure event in the HRMs, as shown in Fig.~\ref{fig5}(h). 
We note that in order to implement quantum communication, Alice measures the qubit and sends the measurement result to Bob as soon as she receives the qubit.

The variances of the GKP qubits are as follows.
In step 1, the sender nodes prepare Bell pairs, where the variances for each constituent qubit are $(\sigma^2,\sigma^2)$, as described in Sec. II C.
Since each mode is transmitted only for a distance of $L_0/2$, the efficiency associated with each loss channel is $\sqrt{\eta}$. For post-, pre-, and CC amplification, the variances after both loss and amplification are $( \sigma^2 + \frac{1-\sqrt{\eta}}{\sqrt{\eta}},\sigma^2+ \frac{1-\sqrt{\eta}}{\sqrt{\eta}})$,  $(\sigma^2+ {1-\sqrt{\eta}},\sigma^2+ {1-\sqrt{\eta}})$, and $( \sigma^2 + \frac{1-\sqrt{\eta}}{2\sqrt{\eta}},\sigma^2+ \frac{1-\sqrt{\eta}}{2\sqrt{\eta}})$, respectively. In the two-way protocol, noise from transmission acts on both the input and the ancilla (in the one-way protocol, it only acts on the input). The variance in the three cases transforms to $( 2\sigma^2 + 2\frac{1-\sqrt{\eta}}{\sqrt{\eta}},2\sigma^2+ 2\frac{1-\sqrt{\eta}}{\sqrt{\eta}})$,  $(2\sigma^2+ {2-2\sqrt{\eta}},2\sigma^2+ {2-2\sqrt{\eta}})$, and $( 2\sigma^2 + \frac{1-\sqrt{\eta}}{\sqrt{\eta}},2\sigma^2+ \frac{1-\sqrt{\eta}}{\sqrt{\eta}})$, respectively, before the application of perfect error correction. After SQEC, the variances of the GKP states are reset to $({\sigma_{q}}^2,{\sigma_{p}}^2)=(\sigma^2,\sigma^2)$ by the finite energy effects of the ancilla states. The process then repeats.

The $X$ and $Z$ logical error probabilities $E^X$ and $E^Z$ for a single transmission step are both independent and equal to
\begin{align}
	E_{\delta}\left(2\sigma^2 + 2\frac{1-\sqrt{\eta}}{\sqrt{\eta}}\right) \quad\quad &\text{two-way post-amplification} \label{eq:twpoa}\\
		E_{\delta}\left(2\sigma^2+ {2-2\sqrt{\eta}}\right) \quad\quad &\text{two-way preamplification} \label{eq:twpra}\\
			E_{\delta}\left(2\sigma^2 + \frac{1-\sqrt{\eta}}{\sqrt{\eta}}\right) \quad\quad &\text{two-way CC-amplification} \label{eq:twcca}	
\end{align}	

\begin{figure}[t]
	\centering \includegraphics[angle=0, scale=1.0]{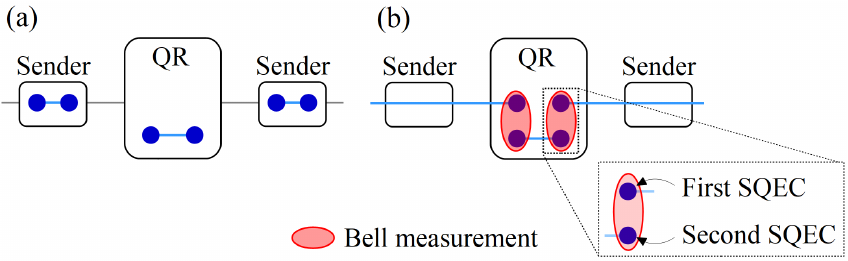} 
	\caption{The protocol with the second SQEC, which is applied to two-way protocols with post- and preamplifications. (a) Each of senders and each quantum repeater station prepare a Bell pair of GKP qubits. (b) Each quantum repeater station receives Bell pair halves from neighboring sender's nodes, and performs the SQECs by Bell measurements.
}
	\label{fig_twoway_sqec}
\end{figure}
We consider one additional type of repeater protocol, which involves implementing an additional round of teleportation-based error correction within each repeater station. This requires the production of an additional Bell pair at each repeater station, as shown in Fig.~\ref{fig_twoway_sqec}. In this case, only half the inputs to each EPR measurement experience noise due to transmission. Due to this asymmetry, we cannot apply the CC-amplification technique. 

The variances of each GKP state now evolve as follows.
In step 1, the sender nodes prepare Bell pairs, as shown in Fig.~\ref{fig_twoway_sqec}(a). The variances for each constituent qubit are $(\sigma^2,\sigma^2)$, as described in Sec. II C.
Since each mode is transmitted only for a distance of $L_0/2$, the efficiency associated with each loss channel is $\sqrt{\eta}$. For post- and preamplification, the variances after both loss and amplification are $( \sigma^2 + \frac{1-\sqrt{\eta}}{\sqrt{\eta}},\sigma^2+ \frac{1-\sqrt{\eta}}{\sqrt{\eta}})$ and $( \sigma^2 +1-\sqrt{\eta},\sigma^2+ 1-\sqrt{\eta})$, respectively. With access to an extra Bell pair at the repeater station that has not been transmitted, we can perform the first SQEC operation while only contributing an additional unit of finite-energy noise, as shown in Fig.~\ref{fig_twoway_sqec}(b). 
Considering that the variances for the extra Bell pair are $(\sigma^2,\sigma^2)$, the variance of the post- and preamplification cases transforms to $( 2\sigma^2 + \frac{1-\sqrt{\eta}}{\sqrt{\eta}},2\sigma^2+ \frac{1-\sqrt{\eta}}{\sqrt{\eta}})$ and $(2\sigma^2+ {1-\sqrt{\eta}},2\sigma^2+ {1-\sqrt{\eta}})$, respectively, before the application of perfect error correction. The error probabilities for the first SQEC in the post- and preamplification cases are given by $E_{\delta}\left(2\sigma^2 +  \frac{1-\sqrt{\eta}}{\sqrt{\eta}}\right)$ and $E_{\delta}\left(2\sigma^2 + 1-\sqrt{\eta}\right)$, respectively.
Next, a second round of teleportation-based error correction is applied using the Bell pair that connects to the next repeater station (which has undergone transmission and amplification). Before the second SQEC is implemented perfectly, the variance of the input in the post- and preamplification cases are $( 2\sigma^2 + \frac{1-\sqrt{\eta}}{\sqrt{\eta}},2\sigma^2+ \frac{1-\sqrt{\eta}}{\sqrt{\eta}})$ and $(2\sigma^2+ {1-\sqrt{\eta}}, 2\sigma^2+ {1-\sqrt{\eta}})$, respectively. The error probabilities for the second SQEC in the post- and preamplification cases are equal to those for the first SQEC.
The output appears at the next repeater station with variance $( \sigma^2 + \frac{1-\sqrt{\eta}}{\sqrt{\eta}},\sigma^2+ \frac{1-\sqrt{\eta}}{\sqrt{\eta}})$ and $( \sigma^2 +1-\sqrt{\eta},\sigma^2+1-\sqrt{\eta})$, respectively, and the process repeats.

\begin{figure}[t]
\centering \includegraphics[angle=0, scale=0.95]{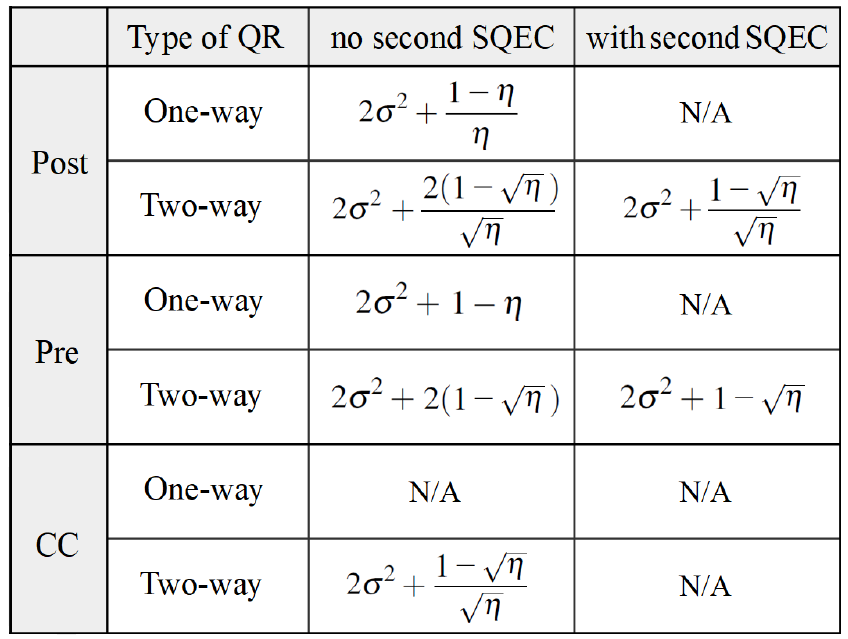} 
\begin{flushleft}
{Table I. Variances used for the calculation of the error probability of the measurement of the GKP qubits. 
}
\end{flushleft}
\label{table}
\end{figure}

Logical $X$ and $Z$ error probabilities $E^X$ and $E^Z$ with the second SQEC are equal to
\begin{align}
	2 E_{\delta}\left(2\sigma^2 + \frac{1-\sqrt{\eta}}{\sqrt{\eta}}\right) \left[ 1- E_{\delta}\left(2\sigma^2 + \frac{1-\sqrt{\eta}}{\sqrt{\eta}}\right) \right] \label{eq:twpoax2}
	\end{align}
for post-amplification, and 
\begin{align}
	2 E_{\delta}\left(2\sigma^2+ {1-\sqrt{\eta}}\right) \left[1- E_{\delta}\left(2\sigma^2+ {1-\sqrt{\eta}}\right) \right] \label{eq:twprax2}
\end{align}	
for preamplification. Note that there are two opportunities per repeater station for each type of logical error to be corrected, one from each round of SQEC.

\begin{figure}[t]
\centering \includegraphics[angle=0, scale=0.9]{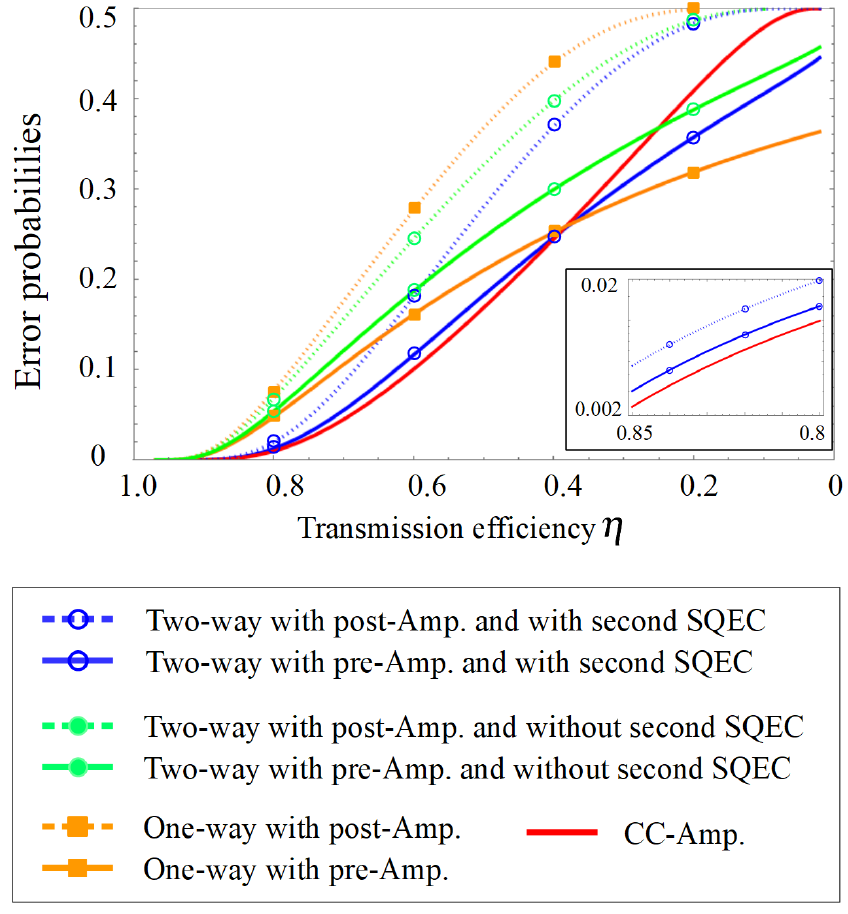} 
\caption{The error probabilities, $E_{\delta}({\sigma_{0}}^2)$, calculated by variances described in Table I, e.g. ${\sigma_{0}}^2 = \sigma^2+(1-\eta)/\eta$ for the one-way protocol without second SQECs, where for brevity the initial variance $\sigma^2$ is set to 0. Note that these error rates are without using the post-selected HRM.}
\label{fig_comparison}
\end{figure}
\subsection{Comparison}

The probabilities $E^{\rm X(Z)}$ of logical $X$ and $Z$ errors are plotted as a function of the transmittance coefficient $\eta$ in Fig.~\ref{fig_comparison}. The minimum error probability for $0.391\leq\eta\leq 1$ is given by the CC-amplification technique. For lower values of $\eta$, one-way preamplification is best. The advantage of the CC-amplification technique is most easy to see by comparing it with the two-way post-amplification-technique without a second round of SQEC in the repeater (i.e., comparing Eq.~(\ref{eq:twcca}) with Eq.~(\ref{eq:twpoa})). Both modify the states after transmission, and both deal with the symmetric noise case. However, the former differs by a factor of $\frac{1}{2}$ in the variance term that arises due to transmission. We attribute this improvement to the use of phase-sensitive amplification instead of phase-insensitive amplification, which introduces additional Gaussian noise. To compare CC-amplification with the one-way techniques, we can consider values of $\eta$ close to 1. For small $1-\eta$ and $\sigma$, the variance before perfect error correction in the one-way protocols is approximately $1-\eta$, whereas for the two-way CC-amplification protocol, it is $(1-\sqrt{\eta})/\sqrt{\eta} \approx (1-\eta)/2$. Thus, the pre- and postamplification one-way protocols are equivalent for small $1-\eta$, but can differ more substantially from the two-way CC-amplification protocol (where we gain a 1/2 factor of excess noise suppression).  
Note that error rates in Fig.~\ref{fig_comparison} are without using the post-selected HRM, i.e., $\delta=0$.

\subsection{Secure key rate}
Here we use the $X$ and $Z$ error probabilities that arise by performing SQEC(s) at each repeater station to determine  secret key rates for the one- and two-way protocols considered above. Given $E^{\rm X}$ and $E^{\rm Z}$ for each repeater station, we can compute the accumulated error rates for the entire transmission from Alice to Bob, denoted $E^{X}_{\rm AB}$ and $E^{Z}_{\rm AB}$ for $X$ and $Z$ errors respectively. 
The secure key rate $R(\geqq0)$ is calculated by \cite{shor2000simple, scarani2009security,muralidharan2014ultrafast},
\begin{equation}
R={P_{\rm suc}}\{1-h(E^{X}_{\rm AB})-h(E^{Z}_{\rm AB})\},
\end{equation}
where $h(E^{X}_{\rm AB})=-E^{X}_{\rm AB}{\rm log}_{2}E^{X}_{\rm AB}-(1-E^{X}_{\rm AB}){\rm log}_{2}(1-E^{X}_{\rm AB})$ and $P_{\rm suc}$ are the binary entropy function and the success probability of each protocol, respectively.

\begin{figure*}[t]
 \centering \includegraphics[angle=0, scale=2.0]{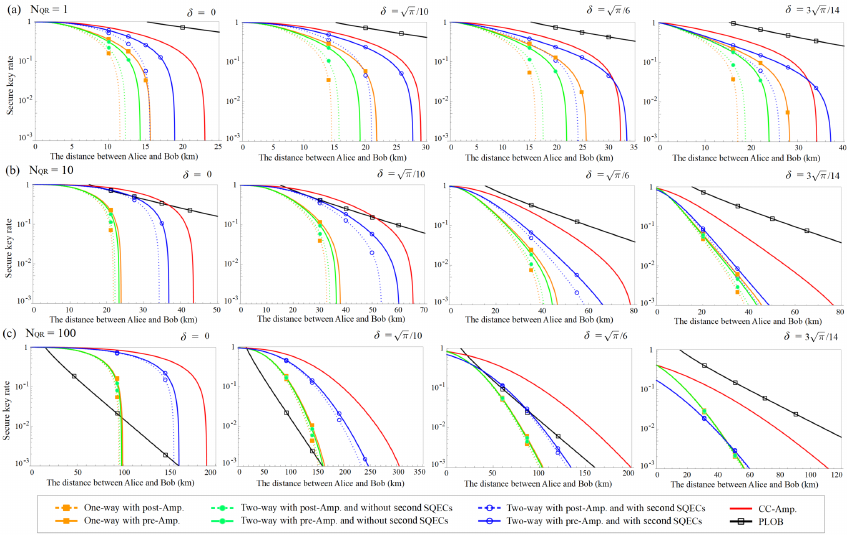} 
\caption{
The secure key rate plotted for the one-way and two-way quantum repeater protocols with three technique of the amplification for the number of repeaters $N_{\rm QR}$=1, 10, and 100, and the values for the HRM $\delta$ =0, $\sqrt{\pi}/10$ ,$\sqrt{\pi}/6$, and $3\sqrt{\pi}/14$. }
\label{fig8}
\end{figure*}

Noting that in all cases we consider, $E^X=E^Z$, the probabilities $E^{X}_{\rm AB}$ and $E^{Z}_{\rm AB}$ can be approximated as 
\begin{equation}
E^{X}_{\rm AB}=E^{Z}_{\rm AB}=\frac{1}{2}\Bigl[ 1-\{1-2E^X\}^{N_{\rm QR}}\Bigr],
\end{equation}
respectively~\cite{muralidharan2014ultrafast}. 
When the HRM is not applied to the protocol, the success probability of every trial of the protocol, $P_{\rm suc}$, is equal to 1, while the HRM with $P_{\rm suc}<1$ could improve the secure key rate due to the reduction of probabilities $E^{\rm Z}$ and $E^{\rm X}$.
$P_{\rm suc}$ with the HRM is given by $(P_{\delta}^{\rm suc})^{2N_{\rm QR}}$ and $(P_{\delta}^{\rm suc})^{4N_{\rm QR}}$ for the protocol without and with second SQECs, respectively, where $P_{\delta}^{\rm suc}$ is the success probability of the HRM in the quantum repeaters as described in Eq. (\ref{eq:psuc}) in Sec. II D.

Figure \ref{fig8} shows the secure key rate plotted as a function of the distance between Alice and Bob for the one- and two-way protocols using the three techniques for amplification, where the distance between Alice and Bob is given by $(N_{\rm QR}+1)L_{0}$, and $L_{0}$ is the distance between neighboring nodes. We assume that the initial squeezing level of the GKP qubit is 15.0 dB, and the transmittance coefficient $\eta$ is equal to ${\rm exp}(-{L_{0}}/{L_{\rm att}})$ with the attenuation length $L_{\rm att}$ = 22 km \cite{ewert2016ultrafast}.
In our numerical calculations, the number of the quantum repeaters is $N_{\rm QR}$=1 and 10, and the values for the HRM $\delta$=0, $\sqrt{\pi}/10$ ,$\sqrt{\pi}/6$, and $\sqrt{\pi}/4$. 
Figure \ref{fig8} shows that the two-way protocol with the CC-amplification is better than the other protocols at performing long-distance quantum communication.
The optimal rate for repeaterless secure key generation is limited by the overall transmission efficiency of direct transmission from Alice to Bob~\cite{takeoka2014fundamental,pirandola2017fundamental}. In Fig.~\ref{fig8}, we  compare secure key rates for our repeaters to the fundamental  Pirandola-Laurenza-Ottaviani-Banchi (PLOB) bound~\cite{pirandola2017fundamental}. The PLOB bound is the best performance achievable in the absence of repeaters and is equal to $- {\rm log}_{2}(1-e^{-{L}/{L_{\rm att}}})$.

\section{Quantum repeater protocol with higher level encoding}\label{Sec4}
In this section, we consider combining GKP error correction with a higher level quantum error correcting code. 
Specifically, we apply our techniques to Varnava's code \cite{varnava2006loss}, which is well suited to dealing with lost qubits.
In Ref.~\cite{azuma2015all}, Varnava's code was used in a scheme for long-distance all-optical quantum communication based on photon qubits. For dual-rail photonic qubits, a lost photon results in a lost qubit heralded by a ``no-click'' event at the detectors. Photon loss does not result in full qubit loss in the case of the GKP qubit encoding. Instead, we artificially introduce a qubit loss whenever the HRM protocol fails (recall that the success probability of the HRM protocol can be decreased to improve the logical error rate of SQEC). This is heralded by the homodyne outcome values of teleportation-based error correction. 
One advantage of using GKP qubits is that entangling operations, such as $C_X$ and $C_Z$ gates, can be implemented deterministically and at room temperature~\cite{yoshikawa2008demonstration, asavanant2019generation, larsen2019deterministic}. These primitives are essential for the implementation of Varnava's code, or other stabilizer codes. 
\begin{figure}[b]
\centering \includegraphics[angle=0, scale=0.8]{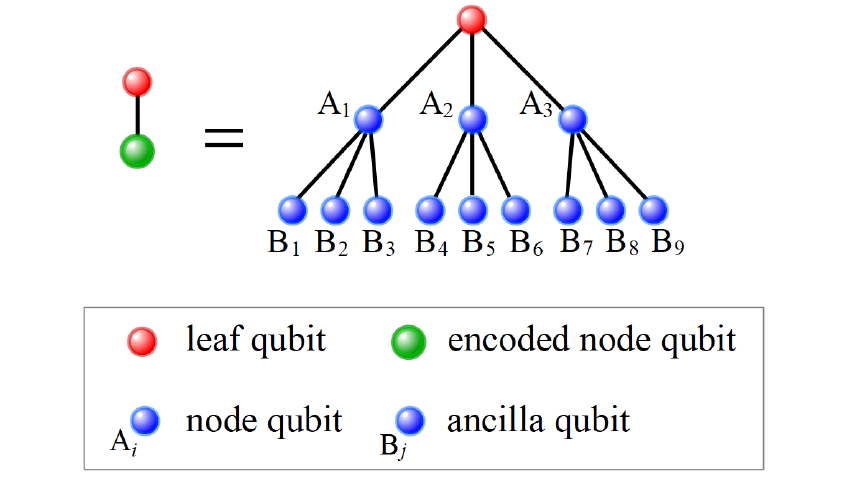}   
 \caption{Cluster state with tree graph, equivalent to an encoded pair between a leaf qubit (red) and an encoded node qubit (green). The encoded node qubit is made up of 3 (blue) node qubits ${\rm A}_{i}$ {$(i=1,2,3)$ and 9 (blue) ancilla qubits ${\rm B}_{3(i-1)+j}$  $(j=1,2,3)$. }
}
\label{varnava}
\end{figure}

\begin{figure*}[t]
\centering \includegraphics[angle=0, scale=2.0]{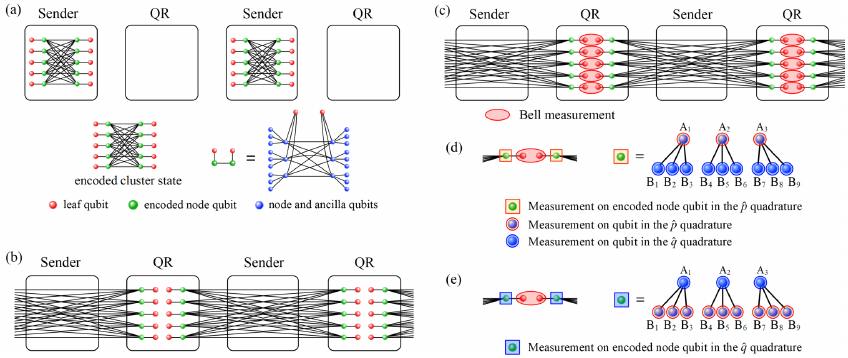} 
    \caption{A schematic drawing of the two-way quantum repeater protocols.
(a) Each sender prepares an encoded cluster state.
(b) Each of repeaters receives the encoded cluster state.
(c) The implementation of the Bell measurement between the leaf qubits.
In the case of using the HRM, the entanglement generation between Alice and Bob succeeds, when the Bell measurements of all groups within quantum repeaters succeeds. 
The Bell measurement of each group succeeds, when at least one of the Bell measurements with the HRM in the group succeeds.
In the case of the path selection, each repeater selects the most reliable result among the Bell measurements in each of groups. Unlike the protocol with the HRM, the path selection realizes deterministic quantum communication.
(d) The $\hat{p}$ measurement on the encoded node qubit.
(e) The $\hat{q}$ measurement on the encoded node qubit.
}
\label{fig10}
\end{figure*}
\subsection{Varnava's code}
Varnava's code is capable of tolerating a qubit loss rate up to $50\%$ \cite{varnava2006loss}.
Figure.~\ref{varnava} shows a cluster state with a tree graph used to implement encoded measurements. 
The cluster state consists of a leaf qubit and an encoded node qubit. The latter is composed of three node qubits and nine ancilla qubits.

First we explain the encoded measurement of the node qubits in the $\hat{q}$ quadrature. 
Let $N_{{\rm A}i} (i=1,2,3) $ be the set of node qubits, ${\rm A}_{i}$, which are connected to the leaf qubit by $C_Z$ gates as shown in Fig.~\ref{varnava}.
The position quadrature of the node qubit ${\rm A}_i$ can be nondestructively measured by measuring the ancilla qubits ${\rm B}_{3(i-1)+j}$ $(j=1,2,3)$ in the $\hat{p}$ quadrature, followed by taking a majority vote of these outcomes. 
This majority voting procedure is equivalent to decoding the three-qubit bit-flip error correcting code, and thus reduces the error of misidentifying the bit value of qubit ${\rm A}_i$ in the $\hat{q}$ quadrature.
This is because the qubits ${\rm A}_i$ and ${\rm B}_{3i+j}$ are stabilized by the operator $\hat{Z}_{{\rm A}i}$ $\hat{X}_{{\rm B}{3(i-1)+j}}$, where ${\hat{Z}}_{{\rm A}i}$ and ${\hat{X}}_{{\rm B}_{3(i-1)+j}}$ are the measurement outcomes of the qubit ${\rm A}_i$ and ${\rm B}_{3(i-1)+j}$ in the $\hat{q}$ and $\hat{p}$ quadrature, respectively. 
Similarly, we can consider the measurement of the three node qubits ${\rm A}_i$ in the $\hat{p}$ quadrature, followed by a majority voting procedure. Measuring node qubit $A_i$ in the $\hat{p}$ quadrature is equivalent to measuring  ${\hat{X}}_{{\rm A}i}\prod_{{\rm B}{3(i-1)+j} \in N_{{\rm A}i}}{\hat{Z}}_{{\rm B}{3(i-1)+j}}$. This procedure is equivalent to implementing and decoding a higher level phase-flip code.

\subsection{Quantum repeater protocol with the HRM}
Now we consider the implementation of Varnava's code with the HRM and CC-amplification.
We describe each step of the proposed method as shown in Fig.~\ref{fig10}, where the proposed method consists of three steps.

In step 1, each sender node in between two quantum repeaters prepares an encoded cluster state, and then sends it to neighboring quantum repeaters, as shown in Fig.~\ref{fig10}(a). 
As an example, we use a cluster state composed of ten leaf qubits and ten encoded node qubits, where the each of encoded node qubits consists of three node qubits and nine ancilla qubits, as shown in Fig.~\ref{fig10}(a).
We note that the encoded cluster state is prepared by using a \emph{fusion gate} along with HRMs. This enables us to construct the cluster state from small-scale entangled states without decreasing the effective squeezing levels (this is explained in more detail in the Appendix). 

In step 2, each of the quantum repeaters receives half the encoded cluster state from neighboring sender's nodes, as shown in Fig.~\ref{fig10}(b). 
Then, repeaters perform the Bell measurements between leaf qubits, as shown in Fig.~\ref{fig10}(c).
In the Bell measurement, repeaters use the HRM to reduce the misidentification of bit values of leaf qubits.
If either of the two leaf qubit measurements within a single EPR measurement fails the HRM procedure, the quantum repeater discards the corresponding link by implementing $\hat{q}$ measurements on the neighboring encoded node qubits. 
Then, if no pair of leaf-qubits succeeds, the current trial of quantum communication is aborted.
When both HRMs in the same Bell measurement succeed, the quantum repeater implements $\hat{p}$ measurements on the neighboring node-qubits to preserve the link. 

Figures.~\ref{fig10}(d) and (e) show the $\hat{p}$ and $\hat{q}$ measurements on the encoded node qubit in the case of Varnava's code with three nodes and three ancilla qubits, respectively.
The $\hat{p}$ measurements on the encoded node qubit are implemented by $\hat{p}$ and $\hat{q}$ measurements on node and ancilla qubits, as shown in Fig.~\ref{fig10}(d).
The logical bit value is then obtained by a majority voting among three joint measurement outcomes, $k_{p,i}^{\rm (n)}\prod k_{q,3(i-1)+j}^{\rm (a)}$ ($i,j$=1,2,3.), where $k_{p,i}^{\rm (n)}$ is the obtained bit value for the $i$-th node qubit, ${\rm A}_{i}$, in the $\hat{p}$ quadrature and $k_{q,3(i-1)+j}^{\rm (a)}$ is the obtained bit value for the $(3(i-1)+j)$-th ancilla qubit, ${\rm B}_{3(i-1)+j}$, in the $\hat{q}$ quadrature.
The $\hat{q}$ measurements on the encoded node qubit are implemented by $\hat{q}$ and $\hat{p}$ measurements on node and ancilla qubits, as shown in Fig.~\ref{fig10}(d).
The bit values of each of node qubits in the $\hat{q}$ quadrature, $k_{q,i}^{\rm (n)}$, are obtained by a majority voting among three measurement outcomes of ancilla qubits in the $\hat{p}$ quadrature, $k_{p,3(i-1)+j}^{\rm (a)}$, respectively.
The logical bit value is then obtained by the bit values of three node qubits as $\sum_{i=1}^{3}k_{q,i}^{\rm (n)}$ mod 2.

To compensate against losses on the node qubits, we rescale the homodyne outcomes by $\sqrt{\eta}^{-1}$. This is equivalent to applying phase-insensitive amplification before the homodyne detectors without rescaling the outcomes, and is analogous to the CC-amplification technique. Loss can then be modeled as a Gaussian-random displacement with variance $(1-\sqrt{\eta})/2\sqrt{\eta}$.
If there are multiple success events, we select the cases with measurement outcomes closest to zero, modulo $\sqrt{\pi}$, as these events are the most reliable. 
In step 3,  Alice and Bob obtain a Bell pair after all feedforward operations (that depend on measurement results from all quantum repeaters).

\begin{figure*}[t]
 \includegraphics[angle=0, scale=2.0]{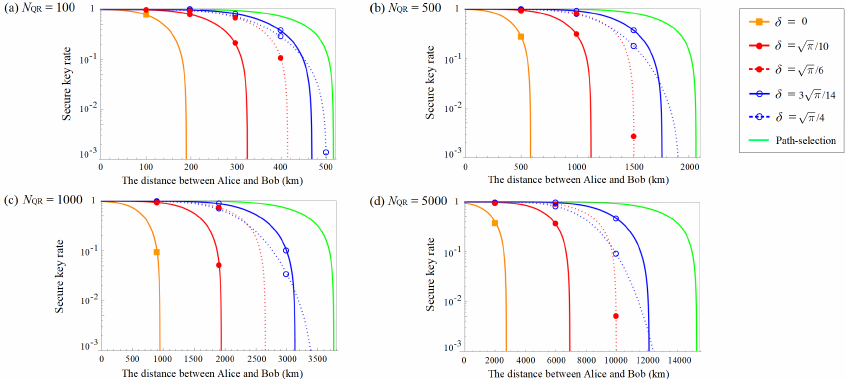} 
\caption{The secure key rate with higher level encoding. $N_{\rm QR}$ is the number of repeaters between Alice and Bob.}
\label{result2}
\end{figure*}

\subsection{Quantum repeater protocol with the path-selection}
We note that the above protocol is not deterministic, since it uses the HRM protocol in implementing the Bell measurement. An alternative deterministic protocol involves always preserving the link with the most reliable measurement results in step 2.
To compare the relative reliability of the different measurement results at each quantum repeater, we employ a Gaussian function---which the true deviation of the GKP qubit $\overline{\Delta}$ obeys---as a likelihood function described as $f(\overline{\Delta})=\frac{1}{\sqrt{2\pi\sigma^{2}}} \mathrm{e}^{-\overline{\Delta}^{2}/(2\sigma^{2})}$.
In the protocol with the path-selection, the likelihood for the Bell measurement between leaf qubits is calculated using $f({\Delta}_{{\rm m},2i-1}) f({\Delta}_{{\rm m},2i})$, where ${\Delta}_{{\rm m},2i-1}$ and ${\Delta}_{{\rm m},2i}$ are measured deviations of the leaf qubits for the $i$ th Bell measurement.
In the case of five leaf qubits, as shown in Fig.~\ref{fig10}, we compare five likelihoods $f({\Delta}_{{\rm m},2i-1}) f({\Delta}_{{\rm m},2i})$ ($i$=1,2,3,4,5), and select the largest likelihood as the most reliable result.
The encoded node qubit attached to leaf qubits resulting in the highest likelihood are measured in $p$, as shown in Fig.~\ref{fig10}, and the other encoded node qubits are measured in $q$, as shown in Fig.~\ref{fig10}(e). We note that if the leaf qubit has bit- or phase-flip errors, we select the incorrect result and an error will occurs in the entanglement generation process between Alice and Bob.

\subsection{Secure key rate}
We characterize this protocol by calculating the secure key rate.
Computing the secure key rate requires us to take into account errors from measuring the leaf qubits, errors in the encoded measurement of the node qubits, and errors during the construction of the encoded cluster.

We begin by describing the error probabilities of the encoded measurements in each quantum repeater, assuming the encoded cluster composed of 10 leaf qubits and 10 encoded node qubits, and that each of the encoded node qubits consists of three node qubits and nine ancilla qubits, as described in Fig.~\ref{result2}.
Since a logical $\hat{q}$ measurement on the node qubit  ${\rm A}_i$ $(i=1,2,3)$ is implemented by measuring the three ancilla qubits ${\rm B}_j$ in the $\hat{p}$ basis, the error probability of each node qubit in the $\hat{q}$ quadrature is calculated by $3(1-e_{{\rm B},p})^2$, where $e_{{\rm B},p}$ is the probability of misidentifying the bit value of the ancilla qubit in the $\hat{p}$ quadrature. 
With three node qubits, the error probability of the encoded $\hat{q}$ quadrature measurement, $E_{\rm en}^{X}$, is 
\begin{equation}
E_{\rm en}^{X}=1-\{1-3(1-e_{{\rm B},p})^2)\}^3,
\end{equation}
to leading order. Then, measuring in the encoded basis of the node qubits ${\rm A}_{i}$ in the $\hat{p}$ quadrature, the measurement result of the logical qubit including the node qubit and ancilla qubit, ${\rm A}_{{\rm L}i}=X_{{\rm A}i}\prod_{{\rm B}j \in N_{{\rm A}i}}Z_{{\rm B}_{3(i-1)+j}}$ $(j=1,2,3)$, and the logical qubit is encoded by the three-qubit bit-flip code.
Thus, the error probability of the encoded measurements in the $\hat{q}$ quadrature, $E_{Z}$, is calculated by
\begin{equation}
E_{\rm en}^{Z}=3\{1-(1-e_{{\rm A},p})(1-e_{{\rm B},q})^3\}^2,
\end{equation}
where $e_{{\rm A},p}$ and $e_{{\rm B},q}$ are the error probabilities of node and ancilla qubits in the $\hat{p}$ and $\hat{q}$ quadratures, respectively.

In addition, we consider the error derived from the construction of the encoded cluster, which is the probability of misidentifying the bit value in the fusion gate with the HRM.
The error derived from the construction of the encoded cluster, $E_{\rm prep}$, is calculated by
\begin{equation}
E_{\rm prep}=34{\times}E_{\delta}(3\sigma^{2}) + 26{\times}E_{\delta}(2\sigma^{2}), 
\end{equation}
where $E_{\delta}(\sigma^2)$ is the error probability of the HRM for the qubit whose variance is $\sigma^2$ (see also the Appendix for the construction of the encoded cluster).

To obtain the rate at which entanglement is generated between Alice and Bob, we calculate the average of the error probability in the $\hat{p}$ ($\hat{q}$) quadrature on each of the quantum repeaters, $E_{\rm QR}^{X(Z)}$.
$E_{\rm QR}^{X(Z)}$ is given by 
\begin{equation}
E_{\rm QR}^{X(Z)}=1-(1-E_{\rm leaf})(1-E_{\rm prep})(1-E_{\rm en}^{X})(1-E_{\rm en}^{Z})^4,
\end{equation}
where $E_{\rm leaf}$ is the error probability of the measurement of the leaf qubit.
Hence, the error probability $E_{\rm AB}^{Z} (E_{\rm AB}^{X})$ which occurs on the entanglement between Alice and Bob is calculated as 
\begin{equation}
E_{\rm AB}^{X}=E_{\rm AB}^{Z} = \frac{1}{2} \{ 1-(1-2E_{\rm QR}^{X})^{N_{\rm QR}} \} ,\\
\end{equation}
where $N_{\rm QR}$ is the number of repeaters.
Then, the secure key rate $R$ is calculated by $R={P_{\rm suc}}(1-h(E_{\rm AB}^{X})-h(E_{\rm AB}^{Z}))$.

Figure~\ref{result2} shows the results of the numerical calculation for the probabilistic protocol with the HRM.
In Fig.~\ref{result2}, the secure key rate between Alice and Bob is plotted as a function of the distance between Alice and Bob, $L_{\rm AB}$, assuming that the encoded cluster state has ten leaf qubits and the initial squeezing level is 15 dB.

Finally, we consider the resource cost in terms of the average of the total number of qubits consumed in the repeater protocol to generate the entanglement between Alice and Bob. To test our method, we compare it with the method based on photon qubits, for example, Azuma's method using Varnava's code~\cite{azuma2015all}.
According to Azuma's method, the average of the total number of photonic qubits is $4.0 \times10^{7}$ ($4.1 \times10^{6}$ ) to generate the entanglement with the error probability, $E_{\rm AB}^{X}$ = $E_{\rm AB}^{Z}$=$3.5(0.89)\%$, for a distance between Alice and Bob of 5000(1000) km. For our method with the path-selection, we obtained the average of the total number of GKP qubits as $2.2 \times10^{5}$ ($4.3 \times10^{4}$ ) with the error probability, $E_{\rm AB}^{X}=E_{\rm AB}^{Z}$ = $3.0(0.76)\%$, for a distance between Alice and Bob 5000(1000) km.
Therefore, our method with GKP qubits can achieve a comparative performance relative to methods based on photonic qubits while using orders-of-magnitude fewer qubits.

\section{Conclusion}\label{conc}
We proposed one- and two-way repeater protocols based on GKP encoded qubits. We have obtained high communication rates for long transmission distances by combining three key innovations: (1) the application of Varnava's code to GKP qubits, (2) the application of highly reliable measurements (HRMs) and path-selection techniques to the code, and (3) the application of the concept of CC amplification, which leverages phase-sensitive amplification on the level of the classical post-processing, thereby circumventing unnecessary additive Gaussian noise. Furthermore, our numerical results showed that these high-communication rates can be achieved with substantially fewer qubits than protocols based on photonic qubits. 

Using the GKP code comes with multiple practical features. Entangling operations and Bell measurements can be implemented deterministically and at room temperature. Furthermore, simple entangled states such as Bell pairs can be generated from product states using passive beamsplitter transformations. Though the implementation of higher level quantum error correcting codes presented here assumed access to a $C_Z$ gate (which involves squeezing operations), we leave to future work the question of whether resources for higher-level quantum error correction can be prepared from GKP product states and passive linear-optical transformations.

\acknowledgments
Recently we became aware of an independent proposal by Rozpedek {\it et al}. for quantum repeaters based on the GKP code~\cite{rozpkedek2021quantum}. The results presented in that work complement our own. It investigates how to combine analog syndrome information available from GKP error correction with simple higher-level qubit codes to better combat noise due to transmission. 

We thank Nicolas Menicucci for useful discussions. K.F. thanks Akihisa Tomita for useful discussions. K.F. acknowledges financial support from JST [Moonshot R$\&$D][Grant No. JPMJMS2064], JST [Moonshot R$\&$D][Grant No. JPMJMS2061], and donations from Nichia Corporation.  R.N.A.\ was supported by National Science Foundation
Award No. PHY-1630114. P.v.L. acknowledges financial support from BMBF via QLinkX and from BMBF/EU-Quantera via ShoQC.

\appendix

\begin{figure*}[htbp]
   \includegraphics[angle=0, width=2\columnwidth]{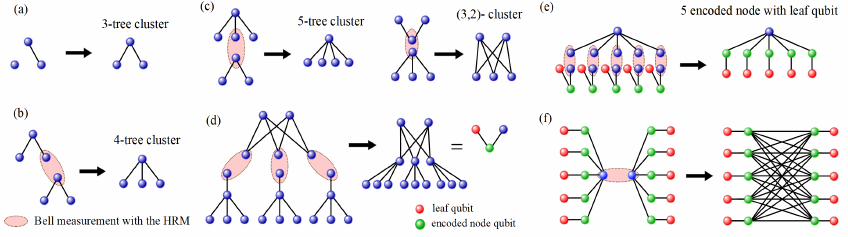} 
\caption{The reliable construction of the encoded cluster state with ten leaf qubits by using the HRM. (a) The preparation of the 3-tree cluster state by using the ${C_Z}$ gate. (b) The generation of the 4-tree cluster state from the two 3-tree cluster states by using the Bell measurement with the HRM. (c) The construction of the 5- and (3,2)-tree cluster states. (d) The construction of the single encoded node qubit with the single leaf qubit.  (e) The construction of five encoded node qubits. (f) The construction of the encoded cluster state.}
\label{construction}
\end{figure*}

\section{Construction of the encoded cluster state with the HRM}
We describe the preparation of the encoded cluster state.
In this work, we use the method introduced in Ref.~\cite{fukui2018high}, where the Bell measurement with the HRM is used to generate the entanglement between two small cluster states to obtain the larger one with a high-reliability.
In Ref.~\cite{fukui2018high}, the Bell measurement is used to prevent the random displacement error of the GKP qubit from propagating during the entanglement generation, and the HRM is used to reduce the error probability of the Bell measurement.

We explain the construction of the encoded cluster state composed of the ten leaf qubits as shown in Fig.~\ref{construction}, where there are four steps. 
In step 1, we prepare the 3-tree cluster composed of a node qubit and two ancilla qubits by using the ${C_Z}$ gate (Fig.~\ref{construction}(a)). 
The $C_Z$ gate, which corresponds to the operator exp(-$i\hat{q}_{\rm C}\hat{q}_{\rm T}$), transforms
\begin{align}
\hat{q}_{\rm C} \to   \hat{q}_{\rm C},\  \hat{p}_{\rm C} \to \hat{p}_{\rm C} - \hat{q}_{\rm T}\ ,   \\
\hat{q}_{\rm T} \to   \hat{q}_{\rm T}, \  \hspace{3pt}  \hat{p}_{\rm T} \to \hat{p}_{\rm T} - \hat{q}_{\rm C}, 
\end{align}
where $\hat{q_{\rm C}}$ ($\hat{q_{\rm T}}$) and $\hat{p_{\rm C}}$ ($\hat{p_{\rm T}}$) are the $\hat{q}$ and $\hat{p}$ quadratures operators of the control (target) qubit, respectively.
We can use the quantum nondemolition gate~\cite{yokoyama2014nonlocal} instead of the $C_{Z}$ gate to generate the entanglement.
Since errors correspond to identically and independently distributed Gaussian random variables, the variance of the control qubit and target qubit in $\hat{p}$ quadrature changes as ${\sigma}^2 \to 2{\sigma}^2$, 
whereas the variance in the $\hat{q}$ quadrature does not change, where we assume that variances of the control and target qubits in both quadratures have the same value, ${\sigma}^2$.

In step 2, we generate the 4-tree cluster state from the two 3-tree cluster states by using the Bell measurement with the HRM (Fig.~\ref{construction}(b)). After the feedforward operation according to the Bell measurement outcome, the 4-tree cluster state is generated. The entanglement generation by using the Bell measurement avoids an increase of the variances of the generated cluster state. In addition, the HRM reduces the error probability of the Bell measurement.
Similarly, the 5- and (3,2)-tree cluster states in Fig.~\ref{construction}(c) are generated by using the Bell measurement with the HRM from the 3- and 4-tree cluster states.

In step 3, we generate the single encoded node qubit with the single leaf qubit from three 5-tree cluster states and the (3,2)-cluster state as shown in Fig.~\ref{construction}(d). Then, we generate the cluster state with five encoded node qubits from five (3,3)-tree cluster states and the 6-tree cluster state (Fig.~\ref{construction}(e)), where 6-tree cluster state is generated from two 4-tree cluster states.
In step 4, we obtain the encoded cluster state from the cluster state with five encoded node qubits (Fig.~\ref{construction}(f)).

\bibliography{ref.bib}

\end{document}